\documentclass{webofc}

\usepackage[varg]{txfonts}   
\usepackage{hyperref}
\usepackage{url}
\usepackage{comment}

\hypersetup{colorlinks=true,citecolor=blue,urlcolor=blue,linkcolor=blue}
\begin{document}
\title{HADES experimental overview}
%
%

\author{\firstname{Hanna} \lastname{Zbroszczyk}\inst{1}\fnsep\thanks{\email{hanna.zbroszczyk@pw.edu.pl}} 
        \firstname{for} \lastname{the HADES Collaboration} 
}

\institute{Warsaw University of Technology, Poland 
          }

\abstract{We report the recent results from the HADES experiment obtained in Au+Au and Ag+Ag collisions at center-of-mass energies per nucleon pair of 2.42 and 2.55 GeV, respectively. In particular, measurements of hadronic and dilepton observables are presented, and prospects for the future experimental program are outlined.

}
\maketitle
\section{Introduction}
\label{intro}
The High Acceptance DiElectron Spectrometer (HADES) is a fixed-target, versatile detector installed at the SIS18 synchrotron at the GSI Helmholtz Center for Heavy-Ion Research in Darmstadt, Germany. HADES operates with heavy-ion beams up to $\sqrt{s_{NN}}$=2.6 GeV, pion beams up to $\sqrt{s}$=2.35 GeV, and proton beams up to $\sqrt{s}$=3.46 GeV. Its broad physics program addresses the microscopic properties of baryon-dominated matter and probes the Equation of State (EoS) through heavy-ion collisions. In addition, HADES performs reference measurements for vacuum and cold QCD matter, and studies the electromagnetic structure of baryons and hyperons. 
In this contribution, we focus on the heavy-ion program, which can reproduce conditions similar to those expected in neutron star mergers \cite{NS1, NS2}.

\section{Physics goals and detector}
\label{goals}

\begin{figure}
\centering
\sidecaption
\includegraphics[width=9cm,clip]{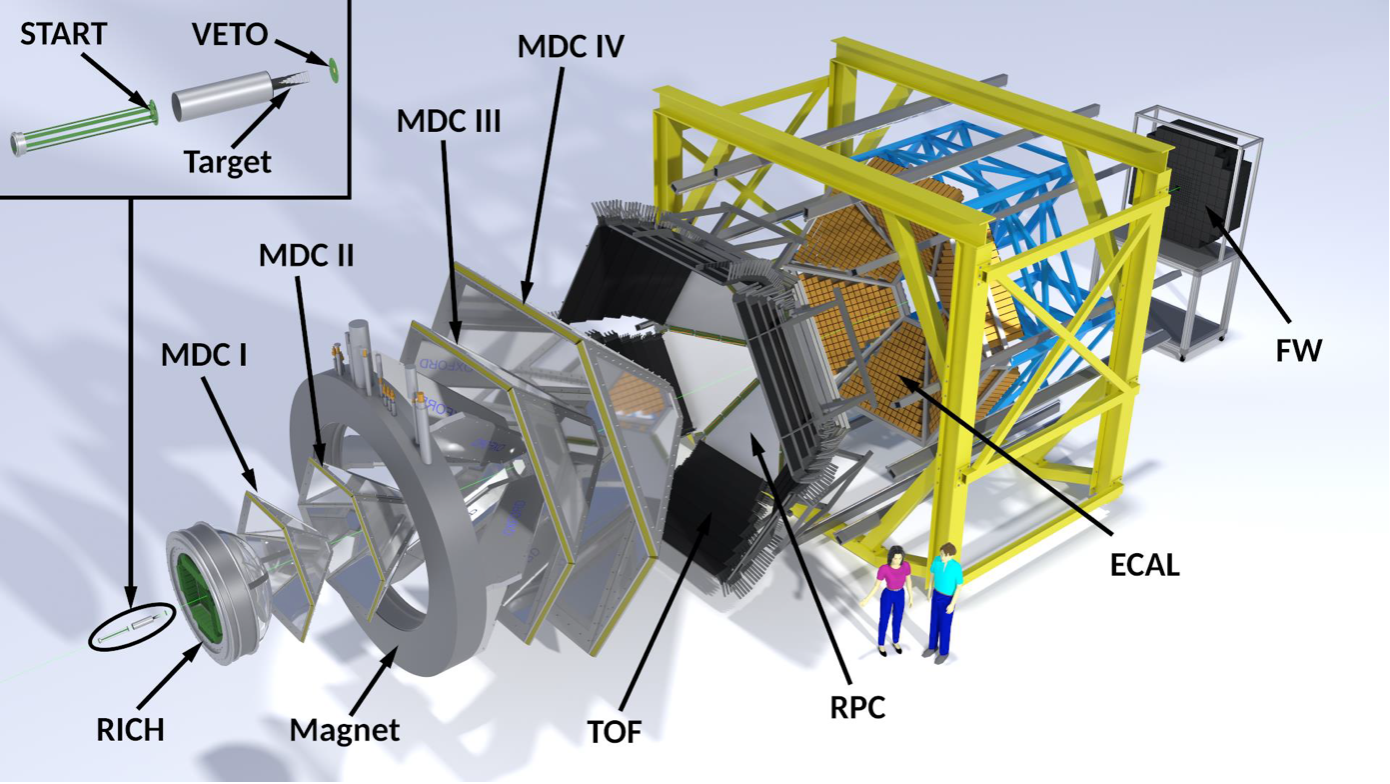}
\caption{Stretched schematic view of the HADES detector with the setup used in heavy-ion collisions focused on the beam-axis direction.}
\label{fig-detector}       
\end{figure}

 HADES \cite{HAD} is a charged-particle detector consisting of a six-coil toroidal magnet surrounding the beam axis and six identical detection sectors located between the coils, providing nearly complete azimuthal coverage. Each sector is equipped with a Ring-Imaging Cherenkov (RICH) detector for dilepton identification, followed by four Multi-Wire Drift Chambers (MDCs) - two upstream and two downstream of the magnetic field - that enable momentum reconstruction and energy-loss measurements. These are complemented by a scintillator  (TOF) and a Resistive Plate Chamber (RPC) for time-of-flight determination. Downstream of the tracking system, an Electromagnetic Calorimeter (ECAL) measures photons, while a forward wall is used for event-plane and centrality determination. In addition, a diamond START detector, located close to the segmented target, provides a precise reaction-time reference. A schematic view of the detector is shown in Fig.~\ref{fig-detector}.

\section{Hadronic Observables}
\label{hadrons}
\subsection{Fluctuations}
\begin{figure}
\centering
\sidecaption
\includegraphics[width=7cm,clip]{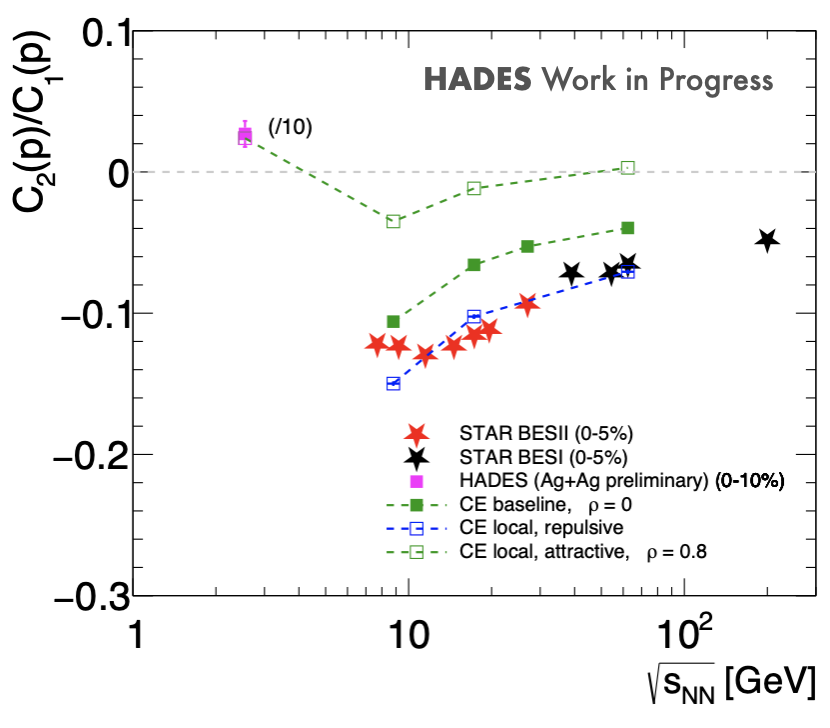}
\caption{The energy dependence of the normalized proton number factorial cumulants $C_{2}(p)/C_{1}(p)$. The purple box represents the preliminary HADES measurement (scaled down by a factor of 10). The model simulations including only global baryon number conservation are shown as solid green squares, while the open green (open blue) boxes correspond to simulations with attractive (repulsive) proton–proton interactions. The open purple box denotes the result of model simulations incorporating three-particle attractions among protons. The latter provides the best description of the HADES data. For details, see~\cite{fl4, marvin}.}
\label{fig-cumulants}       
\end{figure}

Fluctuations of conserved quantities such as electric charge, baryon number, and strangeness are considered sensitive probes of possible phase transitions in strongly interacting matter \cite{fl1, fl2}. These fluctuations should be studied as a function of the baryon chemical potential, which can be accessed by performing heavy-ion collisions at various beam energies. HADES offers a unique opportunity to investigate cumulants and factorial cumulants over a wide rapidity range. In contrast to high-energy measurements, the HADES results on factorial cumulants of proton-number distributions can only be reproduced when, in addition to global baryon number conservation, attractive interactions among protons are taken into account~\cite{fl3,fl4}. This indicates that the transition from predominantly repulsive interactions at high energies to attractive ones occurs at higher baryon chemical potentials, which are probed in heavy-ion collisions at lower energies. The interplay between repulsive and attractive proton–proton interactions, illustrated in Fig.~\ref{fig-cumulants}, provides a consistent explanation for the systematic trends observed across a wide range of collision energies.~\cite{marvin, fl4}. \\
\subsection{Flow}
Azimuthal anisotropy coefficients \cite{flow1, flow2, flow3} of protons and light nuclei are sensitive to the EoS of dense matter \cite{eos}. In addition, the anisotropic behavior of abundantly produced light nuclei at HADES offers an opportunity to study their production mechanisms. Here, in the Figure \ref{fig-hadrons} (left) we report elliptic flow measurements for light nuclei such as deuterons, tritons, and helium-3, studied as a function of transverse momentum. This provides a unique test of coalescence-like scenarios, which successfully describe light nuclei formation.\\
\subsection{Correlations}
\begin{figure*}
\centering
\vspace*{1cm}       
\includegraphics[width=13cm,clip]{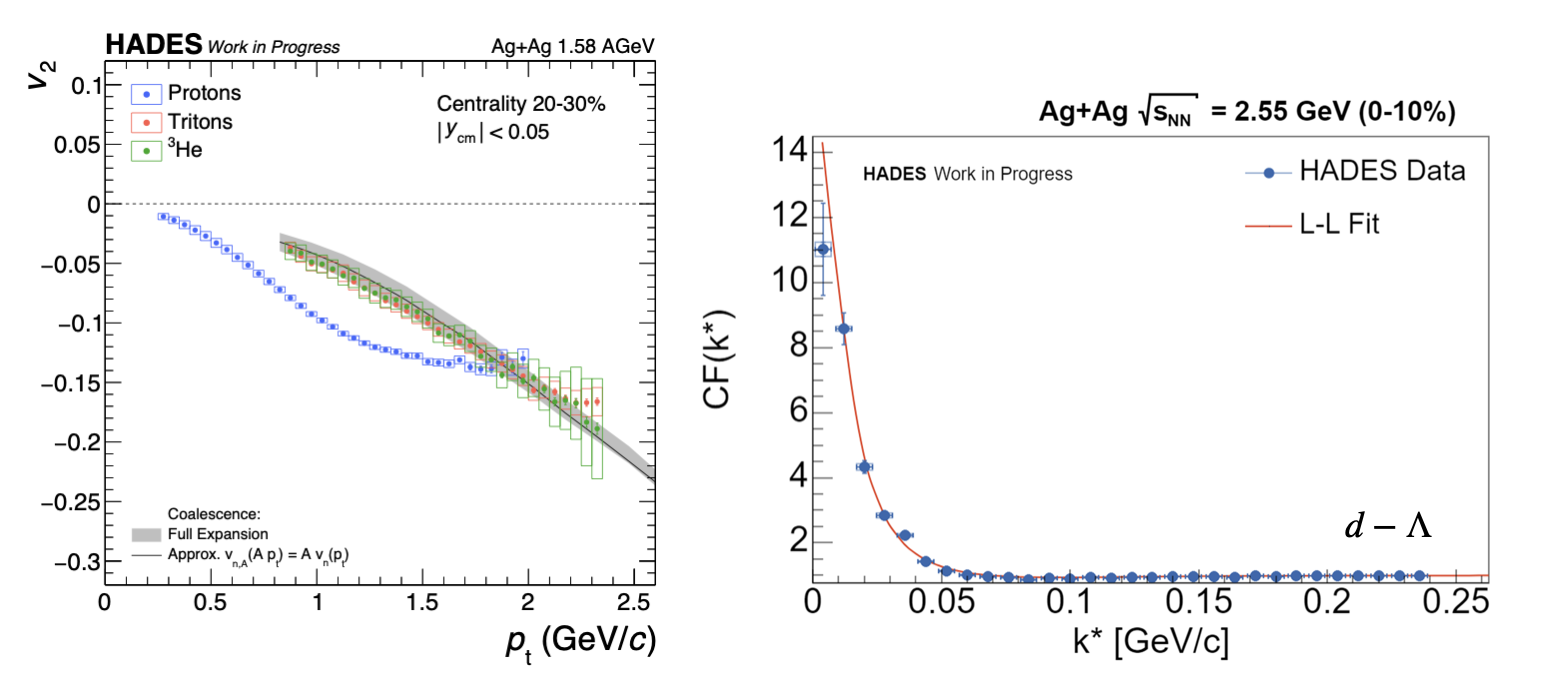}
\caption{Left: scaling of elliptic flow as a function of transverse momentum for protons, tritons and helium-3; right: d-$\Lambda$ femtoscopic correlation function for Ag+Ag collisions together with Lednicky-Lyuboshits fit. }
\label{fig-hadrons}       
\end{figure*}

Interactions between hadrons and light nuclei can be explored not only through collective flow coefficients but also via two-particle correlations arising from quantum statistical effects and/or final-state interactions (Coulomb and strong), accessible through femtoscopy \cite{femto}. Traditionally, femtoscopy assuming well-understood two-particle correlations has been used to probe the space-time characteristics of the particle-emitting source. More recently, with the source geometry parametrized, femtoscopy can be applied to study interactions, particularly those uniquely accessible at HADES energies. Moreover, femtoscopic measurements provide valuable constraints on the preferred EoS for hadronic matter. Figure \ref{fig-hadrons} (right) presents $d-\Lambda$ correlation function for the most central Ag+Ag collisions together with Lednicky-Lyuboshits fit indicating impressive accuracy of the measurement.

\subsection{Hypernuclei}
\label{strangeness}

\begin{figure*}
\centering
\vspace*{1cm}       
\includegraphics[width=13cm,clip]{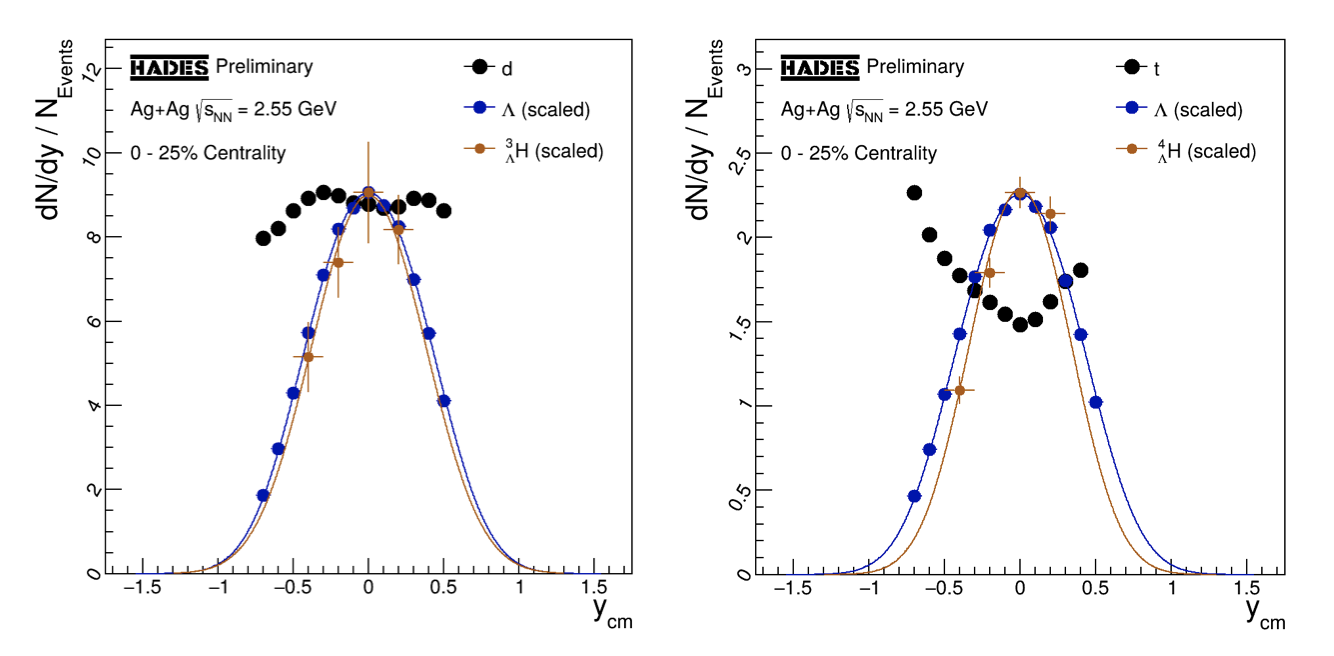}
\caption{Left: preliminary normalized rapidity distributions of deuterons, $\Lambda$ hyperons and hypertritons; right: same but for tritons, $\Lambda$ hyperons and hyperhydrogen-4. }
\label{fig-hipernuclei}       
\end{figure*}

In the HADES energy domain, strangeness production occurs close to the kinematic thresholds in nucleon–nucleon collisions . While light strange hadrons such as $\Lambda$ hyperons and $K^0_S$ mesons can be effectively studied owing to their characteristic decay topologies and relatively long lifetimes, the baryon-rich environment of the fireball also favors the production of hypernuclei. Their yield, however, remains limited by the overall amount of produced strangeness. 
Hypertriton and hyperhydrogen-4 have been reconstructed in Ag+Ag collisions at $\sqrt{s_{NN}}=2.55$ GeV. The available statistics allow for a multi-differential yield analysis as a function of transverse momentum and rapidity. In the same data sample, deuterons, tritons, and $\Lambda$ hyperons have also been reconstructed. This opens the possibility to compare the kinematic distributions of these species in order to investigate whether light hypernuclei exhibit stronger similarities to the distributions of hyperons or to those of light nuclei
Figure~\ref{fig-hipernuclei} displays a comparison of the preliminary $p_{\mathrm{T}}$-integrated rapidity distributions, normalized for comparison, of $\Lambda$ hyperons, deuterons, and hypertriton (left), and of $\Lambda$ hyperons, hyperhydrogen-4, and tritons (right). The distributions of the light hypernuclei clearly exhibit a bell-like shape, closely resembling that of the $\Lambda$ hyperons, while showing a substantial difference compared to the shapes observed for the light nuclei.
\\
Moreover, HADES measures hypertriton and hyperhydrogen-4 lifetimes compatible with the most recent data from STAR, ALICE and J-PARC. 

\section{Dileptons}
\label{dileptons}
Electromagnetic probes provide direct access to all stages of heavy-ion collisions, as they are penetrating observables that do not undergo strong interactions and can traverse the medium without significant alteration \cite{dil1, dil2}. In particular, dilepton measurements allow one to extract both the temperature and the lifetime of the medium.

Nucleon-nucleon reactions serve as an essential baseline for interpreting the heavy-ion data. The measured dilepton signal consists of several distinct contributions: 1) Initial nucleon–nucleon (nn) radiation - originating from first chance p+p, n+n, and p+n collisions,  2) Thermal radiation - emitted from the hot and dense fireball, providing direct access to the temperature of the emitting system 3) Freeze-out cocktail - dilepton pairs from hadronic decays occurring after freeze-out. 

\begin{figure*}
\centering
\vspace*{1cm}       
\includegraphics[width=13cm,clip]{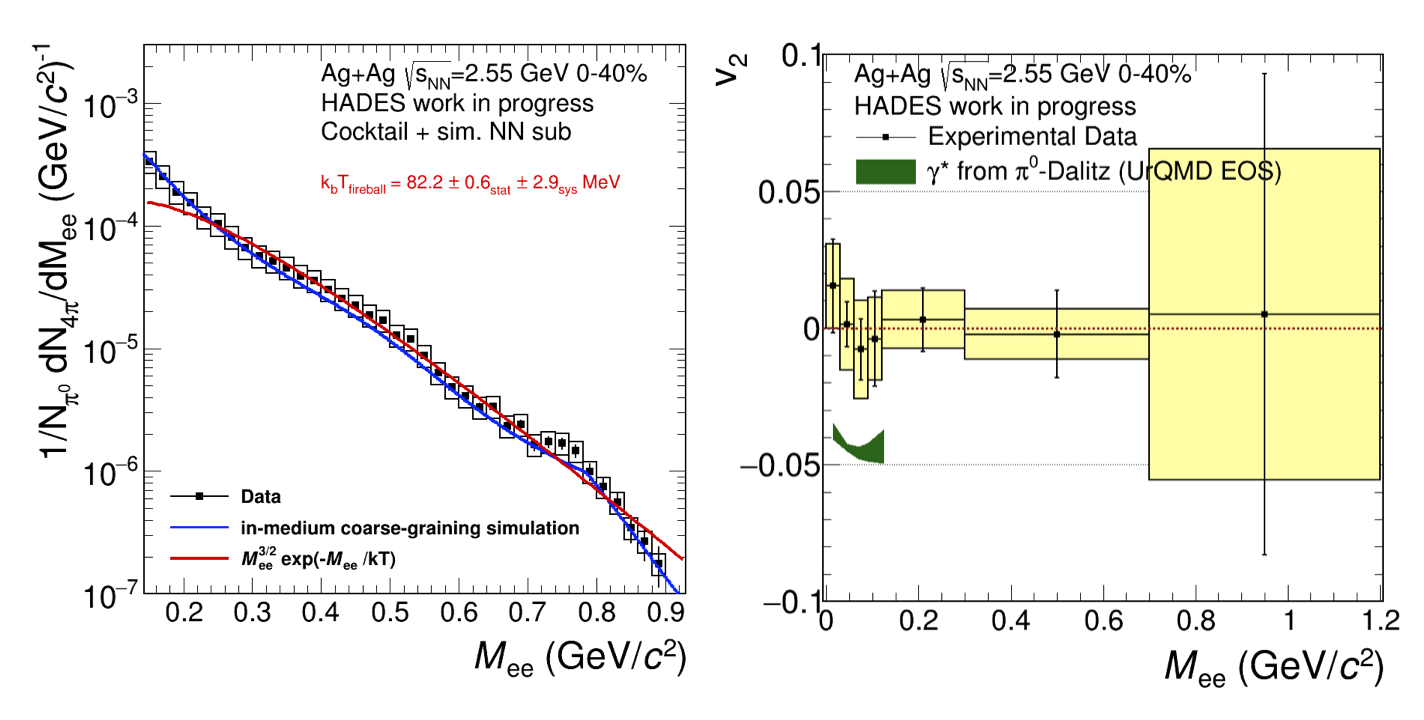}
\caption{Left: Invariant mass of dielectron thermal excess pairs;  right: mass invariant distribution of elliptic flow for dilepton pairs after subtracting contributions originating from $\pi^0$ decays. }
\label{fig-dileptons}       
\end{figure*}

Figure \ref{fig-dileptons} (left) presents the extracted invariant spectra of isolated dielectron pairs, attributed to thermal radiation. The right panel of Fig. \ref{fig-dileptons} shows the elliptic flow of dilepton pairs. 
The thermally dominated dielectron signal shows generally smaller, or even positive, $v_2$ in comparison to the freeze-out sources (e.g. from pions in green). In contrast, a significant difference is observed if compared to the contribution from $\pi^0$ decays (green), which are expected to be emitted on average later in the collision \cite{niklas}. Moreover, it may give insights into the time evolution of the azimuthal anisotropies \cite{dil3}.

\section{Summary and outlook}
\label{summary}
At the Quark Matter 2025 conference, HADES presented results from three key areas: hadrons, strangeness, and dileptons. In the hadron sector, new measurements highlighted the interplay between repulsive and attractive forces among protons, providing a consistent explanation for the observed trends in fluctuation analyses. In the strangeness sector, the focus was on hypernuclei formation mechanisms at high baryon chemical potential. The rapidity distributions of light hypernuclei were found to exhibit a bell-like shape, closely resembling that of the $\Lambda$ hyperons, while differing substantially from those of light nuclei. Finally, dilepton measurements enabled the extraction of both the early-stage temperature and the fireball lifetime. Moreover, HADES observed an elliptic flow consistent with zero for dilepton pairs attributed to thermal radiation.

Looking ahead, in 2025 HADES will continue the energy-scan program for Au+Au collisions in the range of 0.2–0.8 AGeV ($\sqrt{s_{NN}}=1.96$–2.23 GeV), with the aim of searching for critical behavior and probing the limits of the universal freeze-out line. The focus will be on event-by-event particle correlations and fluctuations, dielectron production, strange hadrons, and light nuclei (up to $Z=3$), including flow measurements up to sixth order. In 2026–2027, HADES plans to extend its $\pi$-QCD program to studies of cold matter (in-medium vector mesons, strangeness), hadron spectroscopy, hadron structure and exotics (baryon–meson couplings, electromagnetic couplings, exotic mesons, rare decays), as well as effective interactions (hyperon polarization, hypernuclei formation, hyperon–meson interactions). HADES is expected to remain operational at least until 2030 and, in the coming years, will continue to explore the high–baryon chemical potential region in close synergy with the CBM experiment \cite{cbm}.

\section*{Acknowledgements}
This work work was supported by the Grant of National Science Centre, Poland, No: 2020/38/E/ST2/00019.

%
%

\end{document}